\pdfoutput=1
\documentclass[a4paper]{article}
\usepackage[utf8]{inputenc}

\usepackage{hyperref}
\usepackage{amsmath,amsfonts,amssymb}
\usepackage{bbm}
\usepackage{amsthm}
\usepackage{color}
\usepackage{graphicx}
\usepackage{tikz-cd}
\usepackage{tikz-3dplot}
\usetikzlibrary {decorations.pathmorphing,shadows,fadings}
\usepackage{pgfplots}
\usepackage{cite}

\def\ds{\displaystyle}
\def\bea{\begin{array}{c}}
\def\ea{\end{array}}
\def\be{\begin{equation}\bea\ds}
\def\ee{\ea\end{equation}}
\def\bee{\begin{equation}\begin{array}{rcl}\ds}
\def\eee{\end{array}\end{equation}}

\def\nn{\nonumber}
\def\Lc{{\mathcal{L}}}

\def\Um{{\mathbb{U}}}

\def\Tr{{\rm Tr}\,}

\def\nn{\nonumber}

\title{Jones polynomials from matrix elements of tangles in a pseudounitary representation}
\author{Dmitry Melnikov}
\date{}

\begin{document}

\maketitle

\begin{center}
\textit{\small International Institute of Physics, Federal University of 
Rio Grande do Norte, \\ Campus Universit\'ario, Lagoa Nova, Natal-RN  
59078-970, Brazil}

\vspace{2cm}

\end{center}

\vspace{-2cm}

\begin{abstract}
     In these notes we review the calculation of Jones polynomials using a matrix representation of the braid group and Temperley-Lieb algebra. The pseudounitary representation that we consider allows constructing ``states'' from the group/algebra matrices and compute the knot invariants as matrix elements, rather than traces. In comparison with a more standard way of computing the invariants through traces, the matrix element method is more interesting and complete from the point of view of applications. As a byproduct of the discussion we prove a general formula for pretzel knots. 
\end{abstract}

\bigskip

Since the original discovery of the Jones polynomials~\cite{Jones:1985dw}, many different methods of computing them have been constructed.\footnote{An incomplete list of examples, known to the author, where some of the methods are introduced includes~\cite{Jones:1987dy,Kauffman:1987sta,Witten:1988hf,Turaev:1988eb,Reshetikhin:1990pr,Reshetikhin:1991tc,Kaul:1991np,RamaDevi:1992np,Rosso:1993vn,Lawrence1999wit,Khovanov:1999qla,Ramadevi:2000gq,Morozov:2010kv,Mironov:2011aa,Cherednik:2011nr,Mironov:2013qaa,Mironov:2015aia,Mironov:2015qma}.} One standard approach is based on the theorem of Alexander~\cite{Alexander:1923lem}, stating that any knot can be obtained as tracelike closure of a braid, and the Markov's trace, which guaranties that the trace of a given representation of the braid produces a topological invariant~\cite{Kauffman2001knots}. In these notes we will review a method that can be inferred from papers of Reshetikhin and Turaev~\cite{Reshetikhin:1990pr,Reshetikhin:1991tc}, and from papers of Kauffman et al: Some basic steps of the method also appear in~\cite{Kauffman:2019top,Padmanabhan:2020obt}, for example. This method constructs the Jones polynomials as matrix elements of braids and more general combinations of braids and Temperley-Lieb algebra elements (tangles). Matrix elements of braids are called plats and there is an analogous theorem, stating that any knot (link) can be constructed as a plat closure of a braid~\cite{Birman:1976sta}. Here is an example of a platlike closure of a tangle, for which the method applies:
\be
\label{Whitehead}
\begin{array}{c}\begin{tikzpicture}[baseline=0]
\draw[thick,rounded corners=2] (0.5,-0.1) -- (0.3,-0.1) -- (0.15,0);
\draw[thick,rounded corners=2] (-0.15,0.2) -- (-0.3,0.3) -- (-0.5,0.3);
\draw[thick,rounded corners=2] (-0.5,-0.1) -- (-0.3,-0.1) -- (0.3,0.3) -- (0.5,0.3);
\newcommand{\x}{2}
\draw[thick,rounded corners=2] (0.5,0.3) -- (1.5,0.3);
\draw[thick,rounded corners=2] (0.5+\x,-0.1) -- (0.3+\x,-0.1) -- (0.15+\x,0);
\draw[thick,rounded corners=2] (-0.15+\x,0.2) -- (-0.3+\x,0.3) -- (-0.5+\x,0.3);
\draw[thick,rounded corners=2] (-0.5+\x,-0.1) -- (-0.3+\x,-0.1) -- (0.3+\x,0.3) -- (0.5+\x,0.3);
\newcommand{\y}{-0.8}
\draw[thick,rounded corners=2] (0.5,-0.1+\y) -- (0.3,-0.1+\y) -- (0.15,0+\y);
\draw[thick,rounded corners=2] (-0.15,0.2+\y) -- (-0.3,0.3+\y) -- (-0.5,0.3+\y);
\draw[thick,rounded corners=2] (-0.5,-0.1+\y) -- (-0.3,-0.1+\y) -- (0.3,0.3+\y) -- (0.5,0.3+\y);
\draw[thick,rounded corners=2] (0.5,-0.1+\y) -- (1.5,-0.1+\y);
\draw[thick,rounded corners=2] (0.5+\x,-0.1+\y) -- (0.3+\x,-0.1+\y) -- (0.15+\x,0+\y);
\draw[thick,rounded corners=2] (-0.15+\x,0.2+\y) -- (-0.3+\x,0.3+\y) -- (-0.5+\x,0.3+\y);
\draw[thick,rounded corners=2] (-0.5+\x,-0.1+\y) -- (-0.3+\x,-0.1+\y) -- (0.3+\x,0.3+\y) -- (0.5+\x,0.3+\y);
\renewcommand{\x}{1}
\renewcommand{\y}{-0.4}
\draw[thick,rounded corners=2] (0.5+\x,0.3+\y) -- (0.3+\x,0.3+\y) -- (0.15+\x,0.2+\y);
\draw[thick,rounded corners=2] (-0.15+\x,0.+\y) -- (-0.3+\x,-0.1+\y) -- (-0.5+\x,-0.1+\y);
\draw[thick,rounded corners=2] (-0.5+\x,0.3+\y) -- (-0.3+\x,0.3+\y) -- (0.3+\x,-0.1+\y) -- (0.5+\x,-0.1+\y);
\renewcommand{\x}{-1}
\draw[thick,rounded corners=2] (0.5+\x,-0.1+\y) -- (0.3+\x,-0.1+\y) arc (-90:-270:0.2) -- (0.5+\x,0.3+\y);
\renewcommand{\x}{3}
\draw[thick,rounded corners=2] (-0.5+\x,-0.1+\y) -- (-0.3+\x,-0.1+\y) arc (-90:90:0.2) -- (-0.5+\x,0.3+\y);
\renewcommand{\x}{3}
\renewcommand{\x}{-1}
\draw[thick,rounded corners=2] (-0.5+\x,-0.1+\y) -- (-0.3+\x,-0.1+\y) arc (-90:90:0.2) -- (-0.5+\x,0.3+\y);
\renewcommand{\x}{-2}
\renewcommand{\y}{0}
\draw[thick,rounded corners=2] (0.5+\x,-0.1+\y) -- (0.3+\x,-0.1+\y) arc (-90:-270:0.2) -- (0.5+\x,0.3+\y);
\renewcommand{\y}{-0.8}
\draw[thick,rounded corners=2] (0.5+\x,-0.1+\y) -- (0.3+\x,-0.1+\y) arc (-90:-270:0.2) -- (0.5+\x,0.3+\y);
\renewcommand{\x}{4}
\renewcommand{\y}{0}
\renewcommand{\x}{-2}
\draw[thick,rounded corners=2] (0.5+\x,0.3) -- (1.5+\x,0.3);
\draw[line width=2,gray,dashed,opacity=0.5] (0.5+\x,-1.1) -- (0.5+\x,0.5);
\renewcommand{\x}{+2}
\draw[thick,rounded corners=2] (0.5+\x,0.3) -- (0.7+\x,0.3) arc (90:-90:0.6) -- (0.5+\x,-0.9);
\draw[line width=2,gray,dashed,opacity=0.5] (0.5+\x,-1.1) -- (0.5+\x,0.5);
\renewcommand{\y}{-0.8}
\renewcommand{\x}{-2}
\draw[thick,rounded corners=2] (0.5+\x,-0.1+\y) -- (1.5+\x,-0.1+\y);
\end{tikzpicture}
\end{array}
\ = \ \langle\Psi| T|\Phi\rangle\,. 
\ee

In order to compute the Jones polynomials we will use a specific R matrix representation of the braid group:
\be
\label{Rmatrix}
R \ = \ \left(
\begin{array}{cccc}
    A & & &   \\
     & A-A^{-3} & A^{-1} & \\
     & A^{-1} & 0 & \\
     &  &  & A
\end{array}
\right).
\ee
This R matrix is a well-known solution of the parameter-independent Yang-Baxter equation~\cite{Turaev:1988eb,Jimbo:1985ua}
\be
(R\otimes \mathbbm{1}_2)(\mathbbm{1}_2\otimes R)(R\otimes \mathbbm{1}_2) \ = \ (\mathbbm{1}_2\otimes R)(R\otimes \mathbbm{1}_2)(\mathbbm{1}_2\otimes R)\,,
\ee
where $\mathbbm{1}_2$ is a $2\times 2$ identity matrix. A representation of the Artin's braid group can then be obtained from tensor products of $R$ and $\mathbbm{1}_2$:
\begin{eqnarray}
b_1 & = & R\otimes \mathbbm{1}_2\otimes\cdots \otimes \mathbbm{1}_2\,, \nn \\
b_2 & = & \mathbbm{1}_2\otimes R \otimes\cdots \otimes \mathbbm{1}_2\,, \nn \\
 & \cdots &  \label{braidgens}\\
b_{n-1} & = & \mathbbm{1}_2\otimes\cdots \otimes \mathbbm{1}_2\otimes R\,. \nn
\end{eqnarray}
Here each product contains $n-1$ terms, so that $b_k$ are the generators of the Artin's braid group $B_n$ on $n$ strands.

Given the R matrix one can construct the generators of the Temperley-Lieb algebra. In particular, considering
\be
\label{Udef}
R \ = \ A\, \mathbbm{1}_4 + A^{-1}\, \mathbb{U}\,,
\ee
where
\be
\label{Umat}
\mathbb{U} \ = \ \left(
\begin{array}{cccc}
    0 & & &   \\
     & -A^{2} & 1 & \\
     & 1 & - A^{-2} & \\
     &  &  & 0
\end{array}
\right),
\ee
it is straightforward to check that 
\be
\mathbb{U} ^2 \ = \ d \,\mathbb{U} \,, \qquad d \ = \ -A^2 - A^{-2}\,,
\ee
and 
\begin{eqnarray}
(\mathbb{U} \otimes \mathbbm{1}_2)(\mathbbm{1}_2\otimes \mathbb{U} )(\mathbb{U} \otimes \mathbbm{1}_2) & = & (\mathbb{U} \otimes \mathbbm{1}_2)\,, \nn \\
(\mathbbm{1}_2\otimes \mathbb{U} )(\mathbb{U} \otimes \mathbbm{1}_2)(\mathbbm{1}_2\otimes \mathbb{U} ) & = & (\mathbbm{1}_2\otimes \mathbb{U} )\,,
\end{eqnarray}
which are the defining relations of the Temperley-Lieb algebra. For the Temperley-Lieb algebra $TL_n$ on $n$ strands, one consequently defines the generators as
\begin{eqnarray}
u_1 & = &  \mathbb{U}\otimes \mathbbm{1}_2\otimes\cdots \otimes \mathbbm{1}_2\,, \nn \\
u_2 & = & \mathbbm{1}_2\otimes  \mathbb{U} \otimes\cdots \otimes \mathbbm{1}_2\,, \nn \\
 & \cdots &  \label{TLgens}\\
u_{n-1} & = & \mathbbm{1}_2\otimes\cdots \otimes \mathbbm{1}_2\otimes  \mathbb{U}\,. \nn
\end{eqnarray}

Note that relation~(\ref{Udef}) can be cast in the diagrammatic form, using the standard diagrammatic presentation for the participating matrices:
\be
\label{skein}
\begin{array}{c}
     \begin{tikzpicture}[thick]
     \draw (0,0.5) -- (0.3,0);
     \draw[line width=3,white] (0,0) -- (0.3,0.5);
     \draw (0,0) -- (0.3,0.5);
     \end{tikzpicture}
\end{array}
\ = \ A 
\begin{array}{c}
     \begin{tikzpicture}[thick]
     \draw (0,0) -- (0,0.5);
     \draw (0.3,0) -- (0.3,0.5);
     \end{tikzpicture}
\end{array}
+ A^{-1}
\begin{array}{c}
     \begin{tikzpicture}[thick]
     \draw (0,0) -- (0,0.05) arc (180:0:0.15) -- (0.3,0);
     \draw (0,0.5) -- (0,0.45) arc (180:360:0.15) -- (0.3,0.5);
     \end{tikzpicture}
\end{array}\,.
\ee
This diagrammatic relation is also known as the Conway skein relation in the conventions of Kauffman~\cite{Kauffman:1987sta}. By rotating all diagrams in the skein relation by 90 degrees (note that such a rotation exchanges $\mathbbm{1}_4$ and $\mathbb{U}$), one obtains the skein relation for the the inverse of the R matrix:
\be
\label{iskein}
R^{-1} \ \equiv \ \begin{array}{c}
     \begin{tikzpicture}[thick]
     \draw (0,0) -- (0.3,0.5);
     \draw[line width=3,white] (0,0.5) -- (0.3,0.0);
     \draw (0,0.5) -- (0.3,0);
     \end{tikzpicture}
\end{array}
\ = \ A
\begin{array}{c}
     \begin{tikzpicture}[thick]
     \draw (0,0) -- (0,0.05) arc (180:0:0.15) -- (0.3,0);
     \draw (0,0.5) -- (0,0.45) arc (180:360:0.15) -- (0.3,0.5);
     \end{tikzpicture}
\end{array}
+ A^{-1} 
\begin{array}{c}
     \begin{tikzpicture}[thick]
     \draw (0,0) -- (0,0.5);
     \draw (0.3,0) -- (0.3,0.5);
     \end{tikzpicture}
\end{array}\,.
\ee
It is straighforward to verify that this relation is satisfied in the matrix form using~(\ref{Rmatrix}) and~(\ref{Umat}).

With the explicit form of the braid and Temperley-Lieb generators one can proceed by constructing braids and more general tangles (combinations of braiding and $\mathbb{U}$ operations) necessary to make knots or links, for example, by closure. As a simple example, consider the braid
\be
\begin{array}{c}
     \begin{tikzpicture}[thick]
     \draw (0,0) -- (0.25,0);
     \draw (0,0.5) -- (0.25,0.5);
     \newcommand{\x}{0.25}
     \newcommand{\y}{0}
     \draw[rounded corners=2] (\x,\y) -- (\x+0.25,\y) -- (\x+0.75,\y+0.5) -- (\x+1.,\y+0.5);
     \draw[line width=3,white,rounded corners=2] (\x,\y+0.5) -- (\x+0.25,\y+0.5) -- (\x+0.75,\y) -- (\x+1.,\y);
     \draw[rounded corners=2] (\x,\y+0.5) -- (\x+0.25,\y+0.5) -- (\x+0.75,\y) -- (\x+1.,\y);
     \renewcommand{\x}{1.25}
     \renewcommand{\y}{0}
     \draw[rounded corners=2] (\x,\y) -- (\x+0.25,\y) -- (\x+0.75,\y+0.5) -- (\x+1.,\y+0.5);
     \draw[line width=3,white,rounded corners=2] (\x,\y+0.5) -- (\x+0.25,\y+0.5) -- (\x+0.75,\y) -- (\x+1.,\y);
     \draw[rounded corners=2] (\x,\y+0.5) -- (\x+0.25,\y+0.5) -- (\x+0.75,\y) -- (\x+1.,\y);
     \renewcommand{\x}{2.25}
     \renewcommand{\y}{0}
     \draw[rounded corners=2] (\x,\y) -- (\x+0.25,\y) -- (\x+0.75,\y+0.5) -- (\x+1.,\y+0.5);
     \draw[line width=3,white,rounded corners=2] (\x,\y+0.5) -- (\x+0.25,\y+0.5) -- (\x+0.75,\y) -- (\x+1.,\y);
     \draw[rounded corners=2] (\x,\y+0.5) -- (\x+0.25,\y+0.5) -- (\x+0.75,\y) -- (\x+1.,\y);
     \draw (\x+1,\y) -- (\x+1.25,\y);
     \draw (\x+1,\y+0.5) -- (\x+1.25,\y+0.5);
     \end{tikzpicture}
\end{array}
\ = \ R^3\,,
\ee
whose closure produces the trefoil knot:
\be
\label{trefoil}
\begin{array}{c}
     \begin{tikzpicture}[thick]
     \newcommand{\x}{0}
     \newcommand{\y}{0}
     \draw (\x,\y) arc (90:270:0.25);
     \draw (\x,\y+0.5) arc (90:270:0.75);
     \draw (0,0) -- (0.25,0);
     \draw (0,0.5) -- (0.25,0.5);
     \renewcommand{\x}{0.25}
     \renewcommand{\y}{0}
     \draw[rounded corners=2] (\x,\y) -- (\x+0.25,\y) -- (\x+0.75,\y+0.5) -- (\x+1.,\y+0.5);
     \draw[line width=3,white,rounded corners=2] (\x,\y+0.5) -- (\x+0.25,\y+0.5) -- (\x+0.75,\y) -- (\x+1.,\y);
     \draw[rounded corners=2] (\x,\y+0.5) -- (\x+0.25,\y+0.5) -- (\x+0.75,\y) -- (\x+1.,\y);
     \renewcommand{\x}{1.25}
     \renewcommand{\y}{0}
     \draw[rounded corners=2] (\x,\y) -- (\x+0.25,\y) -- (\x+0.75,\y+0.5) -- (\x+1.,\y+0.5);
     \draw[line width=3,white,rounded corners=2] (\x,\y+0.5) -- (\x+0.25,\y+0.5) -- (\x+0.75,\y) -- (\x+1.,\y);
     \draw[rounded corners=2] (\x,\y+0.5) -- (\x+0.25,\y+0.5) -- (\x+0.75,\y) -- (\x+1.,\y);
     \renewcommand{\x}{2.25}
     \renewcommand{\y}{0}
     \draw[rounded corners=2] (\x,\y) -- (\x+0.25,\y) -- (\x+0.75,\y+0.5) -- (\x+1.,\y+0.5);
     \draw[line width=3,white,rounded corners=2] (\x,\y+0.5) -- (\x+0.25,\y+0.5) -- (\x+0.75,\y) -- (\x+1.,\y);
     \draw[rounded corners=2] (\x,\y+0.5) -- (\x+0.25,\y+0.5) -- (\x+0.75,\y) -- (\x+1.,\y);
     \renewcommand{\x}{3.25}
     \renewcommand{\y}{0}
     \draw (\x,\y) -- (\x+0.25,\y);
     \draw (\x,\y+0.5) -- (\x+0.25,\y+0.5);
     \renewcommand{\x}{3.5}
     \renewcommand{\y}{0}
     \draw (\x,\y) arc (90:-90:0.25);
     \draw (\x,\y+0.5) arc (90:-90:0.75);
     \draw (0,\y-0.5) -- (\x,\y-0.5);
     \draw (0,\y-1) -- (\x,\y-1);
     \end{tikzpicture}
\end{array}
\ \sim \ 
\begin{array}{c}
     \includegraphics[height=1.5cm]{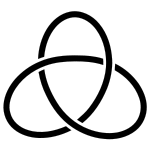} 
\end{array} \,.
\ee

Computing the regular trace of $R^3$ does not give a topological invariant, as expected, and one should instead, use a modified definition. Let us first define~\cite{Turaev:1988eb}
\be
H \ = \ \left(
\begin{array}{cc}
 -A^2 &   \\
     & -A^{-2}  \\
\end{array}
\right).
\ee
This matrix acts in the two-dimensional space $V$ identified with a single strand. To compute the trace of an $n$-strand braid or tangle, we will need the $n$-th tensor power of $H$ acting on $V^{\otimes n}$. In particular, 
\be
\Tr_{\rm M}T^{(n)} \ \equiv \ \Tr \left(H^{\otimes n}T^{(n)}\right)
\ee
defines a Markov's trace of $n$-strand tangle $T^{(n)}$. As a Markov's trace, $\Tr_{\rm M}$ produces a topological invariant. In the trefoil example above
\be
\label{Ktrefoil}
\Tr_{\rm M}R^3 \ = \ (-A^{3})^3 d\left(-A^{-16}+A^{-12}+A^{-4}\right).
\ee
Modulo factor $d$ this is the bracket polynomial of Kauffman, which is related to the Jones polynomial by extracting the prefactor $(-A^3)^{w}$, where $w$ is called writhe of the knot, computed as a sum of the signs of the crossings in the knot diagram. The bracket polynomial is invariant under planar, but not ambient isotopies (first Reidemeister move), which is corrected by the writhe factor. Dividing by $(-A^3)^{w}$ (with $w=3$) and replacing $A^{-4}\to q$, gives the Jones polynomial $-q^{4}+q^{3}+q^{1}$ of the trefoil, invariant under the ambient space isotopies.\footnote{Note that $q\to 1/q$ ($A\to 1/A$) relates the invariants of a knot and of its mirror image. The Jones polynomials are defined with this ambiguity in the conventions.}

The bracket polynomial of a link diagram $\langle \Lc \rangle$ can be computed by using a simple algorithm based on skein relations~(\ref{skein}) and~(\ref{iskein})~\cite{Kauffman:1987sta}: 
\begin{enumerate}
    \item If $\Lc$ is a trivial unknotted and unlinked circle then the bracket evaluates to unity:
    \be
    \label{bracket0}
    \langle \begin{array}{c}
         \begin{tikzpicture}
             \draw[thick] (0,0) circle (0.2cm);
         \end{tikzpicture} 
    \end{array}
    \rangle = 1\,.
    \ee
   \item If a link is has an unknotted unlinked circle as its component than the bracket factorizes into the product of the number $d$ and the bracket of the remaining component:
   \be
   \label{bracket2}
   \langle \Lc \rangle = \langle \begin{array}{c}
         \begin{tikzpicture}
             \draw[thick] (0,0) circle (0.2cm);
         \end{tikzpicture} 
    \end{array} \sqcup\ \Lc'\ \rangle \ = \ d\cdot \langle \Lc'\rangle\,.
    \ee 
    \item If the diagram of a link contains a crossing, the bracket of this diagram is ``resolved'' as a sum of two brackets, in which the crossing is replaced by the two respective routings of lines, according to the skein relations~(\ref{skein}) and~(\ref{iskein}):
    \be
\label{bracket3}
\Big\langle \begin{array}{c}
     \begin{tikzpicture}[thick]
     \draw (0,0.5) -- (0.3,0);
     \draw[line width=3,white] (0,0) -- (0.3,0.5);
     \draw (0,0) -- (0.3,0.5);
     \draw[line width=5,gray] (0.15,0.25) circle (0.25);
     \end{tikzpicture}
\end{array}\Big\rangle
\ = \ A 
\Big\langle\begin{array}{c}
     \begin{tikzpicture}[thick]
     \draw (0.07,0) -- (0.07,0.5);
     \draw (0.23,0) -- (0.23,0.5);
     \draw[line width=5,gray] (0.15,0.25) circle (0.25);
     \end{tikzpicture}
\end{array}
\Big\rangle
+ A^{-1}
\Big\langle
\begin{array}{c}
     \begin{tikzpicture}[thick]
     \draw (0,0) -- (0,0.05) arc (180:0:0.15) -- (0.3,0);
     \draw (0,0.5) -- (0,0.45) arc (180:360:0.15) -- (0.3,0.5);
     \draw[line width=5,gray] (0.15,0.25) circle (0.25);
     \end{tikzpicture}
\end{array}
\Big\rangle
\,.
\ee
\end{enumerate}

Meanwhile, in the matrix representation presented here, there will always be an extra factor of $d$, as compared to the standard definitions of Jones and Kauffman. This is related to the normalization of the simplest polynomial, the invariant of the unknot -- the circle, cf. equation~(\ref{bracket0}). Note that the circle is similar to the trace of an identity operator, which computes the dimension of space $V$, but unlike the ordinary trace, Markov's trace gives for the identity
\be
\Tr_{\rm M}\mathbbm{1}_2 \ = \ - A^2 - A^{-2} \ = \ d\,.
\ee
In the context of quantum groups parameter $d$ is called \emph{quantum dimension}. Indeed, we will require that $A$ (and $q$) is a root of unity, so
\be
d \ = \ -A^2-A^{-2} \ = \ -q^{1/2} - q^{-1/2} \ = \ -\frac{q-q^{-1}}{q^{1/2}-q^{-1/2}} \ = \ -[2]\,, 
\ee
where 
\be
[n] \ = \ \frac{q^{n/2}-q^{-n/2}}{q^{1/2}-q^{-1/2}}\,,
\ee
is called quantum number $n$. For affine algebras quantum dimensions, in terms of quantum numbers, are obtained from an extension of the Weyl formula. Here, the Markov's trace gives $[2]$ as the quantum dimension of the spin $1/2$ representation of the $su(2)_k$ algebra, up to the sign. Hence it makes sense to normalize the unknot to quantum dimension $d$. In the Jones' convention it is simply unity. We will thus use the following modification of the rule 1 above~(\ref{bracket0}):
\begin{enumerate}
    \item[1$^\ast$] If $\Lc$ is a trivial unknotted and unlinked circle then the bracket evaluates to the number $d$:
    \be
    \label{bracket1}
    \langle \begin{array}{c}
         \begin{tikzpicture}
             \draw[thick] (0,0) circle (0.2cm);
         \end{tikzpicture} 
    \end{array}
    \rangle = d\,.
    \ee
\end{enumerate}

Now, let us observe that the properties of $\Um$, especially its diagrammatic presentation, suggest that
\be
\label{Ufactor}
\Um \ \equiv \ \begin{array}{c}
     \begin{tikzpicture}[thick]
     \draw (0,0) -- (0.05,0) arc (-90:90:0.15) -- (0,0.3);
     \draw (0.5,0) -- (0.45,0) arc (-90:-270:0.15) -- (0.5,0.3);
     \end{tikzpicture}
\end{array}\ = \ |u\rangle\langle u|\,, \qquad \text{for some} \qquad |u\rangle \in V\otimes V\,.
\ee
In fact, if we were solving a similar problem $\Um = |u\rangle\otimes |u\rangle$, we would immediately find (cf.~\cite{Kauffman:2019top,Padmanabhan:2020obt})
\be
|u\rangle \ = \ \pm (0,iA,-iA^{-1},0)\,.
\ee
Shortly we will explain that since the representation we consider is pseudounitary, representations of $\Um$ as a matrix in $V^{\otimes 2}$ and as a vector in $V^{\otimes 4}$ are equivalent. 

Obviously, we want the following diagrammatic presentation for the vectors:
\be
\label{uvectors}
|u\rangle \ = \ 
\begin{array}{c}
\begin{tikzpicture}[thick]
    \draw (0,0) -- (0.2,0) arc (-90:90:0.25) -- (0,0.5);
\end{tikzpicture}
\end{array}\,,
\qquad 
\langle u| \ = \ 
\begin{array}{c}
\begin{tikzpicture}[thick]
    \draw (0,0) -- (-0.2,0) arc (-90:-270:0.25) -- (0,0.5);
\end{tikzpicture}
\end{array}\,,
\qquad 
|u\rangle\otimes |u\rangle \ = \ 
\begin{array}{c}
\begin{tikzpicture}[thick]
    \draw (0,0) -- (0.2,0) arc (-90:90:0.25) -- (0,0.5);
    \draw (0,1) -- (0.2,1) arc (-90:90:0.25) -- (0,1.5);
\end{tikzpicture}
\end{array}\,,
\qquad \ldots
\ee
The consistency can be checked by a number of examples. Let us here compute the trefoil knot using its plat diagram,
\be
\label{Ktrefoil2}
\begin{array}{c}
     \begin{tikzpicture}[thick]
     \newcommand{\x}{0}
     \newcommand{\y}{0}
     \draw (\x,\y-0.5) -- (\x-0.2,\y-0.5) arc (-90:-270:0.25) -- (\x,\y);
    \draw (\x,\y+0.5) -- (\x-0.2,\y+0.5) arc (-90:-270:0.25) -- (\x,\y+1);
     \draw (0,0) -- (0.25,0);
     \draw (0,0.5) -- (0.25,0.5);
     \renewcommand{\x}{0.25}
     \renewcommand{\y}{0}
     \draw[rounded corners=2] (\x,\y) -- (\x+0.25,\y) -- (\x+0.75,\y+0.5) -- (\x+1.,\y+0.5);
     \draw[line width=3,white,rounded corners=2] (\x,\y+0.5) -- (\x+0.25,\y+0.5) -- (\x+0.75,\y) -- (\x+1.,\y);
     \draw[rounded corners=2] (\x,\y+0.5) -- (\x+0.25,\y+0.5) -- (\x+0.75,\y) -- (\x+1.,\y);
     \renewcommand{\x}{1.25}
     \renewcommand{\y}{0}
     \draw[rounded corners=2] (\x,\y) -- (\x+0.25,\y) -- (\x+0.75,\y+0.5) -- (\x+1.,\y+0.5);
     \draw[line width=3,white,rounded corners=2] (\x,\y+0.5) -- (\x+0.25,\y+0.5) -- (\x+0.75,\y) -- (\x+1.,\y);
     \draw[rounded corners=2] (\x,\y+0.5) -- (\x+0.25,\y+0.5) -- (\x+0.75,\y) -- (\x+1.,\y);
     \renewcommand{\x}{2.25}
     \renewcommand{\y}{0}
     \draw[rounded corners=2] (\x,\y) -- (\x+0.25,\y) -- (\x+0.75,\y+0.5) -- (\x+1.,\y+0.5);
     \draw[line width=3,white,rounded corners=2] (\x,\y+0.5) -- (\x+0.25,\y+0.5) -- (\x+0.75,\y) -- (\x+1.,\y);
     \draw[rounded corners=2] (\x,\y+0.5) -- (\x+0.25,\y+0.5) -- (\x+0.75,\y) -- (\x+1.,\y);
     \renewcommand{\x}{3.25}
     \renewcommand{\y}{0}
     \draw (\x,\y) -- (\x+0.25,\y);
     \draw (\x,\y+0.5) -- (\x+0.25,\y+0.5);
     \renewcommand{\x}{3.5}
     \renewcommand{\y}{0}
     \draw (0,\y-0.5) -- (\x,\y-0.5);
     \draw (0,\y+1) -- (\x,\y+1);
     \draw (\x,\y-0.5) -- (\x+0.2,\y-0.5) arc (-90:90:0.25) -- (\x,\y);
    \draw (\x,\y+0.5) -- (\x+0.2,\y+0.5) arc (-90:90:0.25) -- (\x,\y+1);
     \end{tikzpicture}
\end{array}
\ = \ \langle u\otimes u| R^3 |u\otimes u\rangle \ = \ (-A^{3})^3 d\left(-A^{-16}+A^{-12}+A^{-4}\right).
\ee
This is precisely the result of~(\ref{Ktrefoil}). Here, instead of computing a trace of the braid matrix, we compute its matrix elements with vectors~(\ref{uvectors}). In order to get things right in the given representation we use transposition of vectors rather than their Hermitian conjugation.

Let us discuss the last point in more detail. Since the invariant of the unknot is $d$, we expect that
\be
\label{unorm}
\langle u|u\rangle \ = \ \begin{array}{c}
\begin{tikzpicture}[thick]
    \draw (0,0) -- (0.2,0) arc (-90:90:0.25) -- (0,0.5) arc (90:270:0.25);
\end{tikzpicture}
\end{array} \ = \ d\,.
\ee
Recall that we need $A$ to be a root of unity. Obviously, if we try to compute $(|u\rangle)^\dagger |u\rangle$ the result would be $2$, and not $[2]$. This is a direct consequence of our representation being not unitary, in particular, $R^\dagger \neq R^{-1}$. However, one can check that
\be
(\Sigma\otimes\Sigma)R^\dagger(\Sigma\otimes\Sigma)^\dagger \ = \ R^{-1}\,,
\ee
where
\be
\Sigma \ = \ 
\left(
\begin{array}{cc}
     & 1 \\
    1 & 
\end{array}
\right).
\ee
Consequently, if $\langle u|$, or any other state in the dual space $V^\ast$, is defined as 
\be
\langle u| \ \equiv \ (|u\rangle)^\dagger \cdot\Sigma\,,
\ee
one will get overlaps and matrix elements compatible with the bracket polynomials, such as~(\ref{unorm}) and~(\ref{Ktrefoil2}). This definition also implies $|u\rangle\langle u|=|u\rangle\otimes |u\rangle$, as proposed earlier.

Using braid and Temperley-Lieb generators one can construct other useful diagrammatic states. Let us define
\be
\label{e0}
|e_0\rangle \ = \ |u\rangle \otimes |u\rangle\,.
\ee
Note that,
\be
\label{e1}
|e_1\rangle \ \equiv \ \begin{array}{c}
\begin{tikzpicture}[thick]
    \draw (0,0) -- (0.2,0) arc (-90:90:0.25) -- (0,0.5);
    \draw (0,-0.5) -- (0.2,-0.5) arc (-90:90:0.75) -- (0,1.);
\end{tikzpicture}
\end{array} \ = \ u_2|e_0\rangle \ = \ b_1b_2 |e_0\rangle\,.
\ee
The components of this 16-dimensional vector can be easily computed through the use of the generator matrices~(\ref{braidgens}) and~(\ref{TLgens}). We can use $|e_1\rangle$ for an alternative derivation of the closure~(\ref{trefoil}), in this case as a matrix element
\be
\langle e_1|b_1^3|e_1\rangle \ = \ (-A^{3})^3 d\left(-A^{-16}+A^{-12}+A^{-4}\right), 
\ee
in agreement with the previous results.

As another example, let us compute the bracket polynomial of~(\ref{Whitehead}), which is known as the Whitehead link. One obtains
\be
\langle e_0|u_2b_1^{-1}b_3^{-1}b_2b_3^{-1}b_1^{-1}|e_1\rangle \ = \ (-{A^3})^{-1}d\left(A^{14}-2A^{10} + A^6 - 2 A^2 + A^{-2} - A^{-6}\right).
\ee
Substituting $A^{-4}\to q$ in the last bracket gives the Jones polynomial of the Whitehead link.

As the last example, we compute the bracket polynomial of the pretzel knots, for which a general formula was proposed in~\cite{Galakhov:2014sha} (see also~\cite{Mironov:2014eza,Galakhov:2015fna}). Pretzel knots are the knots that can be drawn on a genus $g$ surface without self-intersections. Let us consider the case of genus $g=3$:
\be
\label{pretzel}
\begin{array}{c}
     \begin{tikzpicture}[thick]
         \foreach \a in {0.,0.8,...,3.2}
            \draw (0,\a) -- (2.5,\a);
         \foreach \a in {0.3,1.1,...,3.5}
            \draw (0,\a) -- (2.5,\a);
         \foreach \a in {0.3,1.1,...,2.7}
            \draw[rounded corners=3] (0,\a) -- (-0.2,\a) -- (-0.2,\a+0.5) -- (0,\a+0.5);
         \foreach \a in {0.3,1.1,...,2.7}
            \draw[rounded corners=3] (2.5,\a) -- (2.5+0.2,\a) -- (2.5+0.2,\a+0.5) -- (2.5,\a+0.5);
        \draw[rounded corners=3] (0,0) -- (-0.5,0) -- (-0.5,2.7) -- (0,2.7);
         \draw[rounded corners=3] (2.5,0) -- (3,0) -- (3,2.7) -- (2.5,2.7);
         \foreach \a in {0.,0.8,...,3.2}
            \fill[gray,rounded corners=2] (0.75,\a-0.2) rectangle (1.75,\a+0.5);
         \foreach \a in {0.,0.8,...,3.2}   
            \fill[white,rounded corners=2,opacity=0.7] (0.85,\a-0.1) rectangle (1.65,\a+0.4);
        \foreach \n in {0,1,...,3}
        \draw (1.25,2.55-\n*0.8) node {$n_\n$};
     \end{tikzpicture} 
\end{array}\,.
\ee
Here the boxes with integer parameters $n_i$ denote insertions of $R^{n_i}$. Let us compute the result for a specific choice of the parameter vector $\{n_i\}=\{1,2,3,4\}$.

The result is computed by
\be
\langle\Psi | R^{n_0}\otimes R^{n_1}\otimes R^{n_2}\otimes R^{n_3}|\Psi\rangle\,,
\ee
where the initial and the final states are prepared via 
\be
|\Psi\rangle \ = \ b_1b_2b_3b_4b_5b_6|u\otimes u\otimes u\otimes u\rangle
\ = \ u_2u_4u_6|u\otimes u\otimes u\otimes u\rangle. 
\ee
The result is 
\be
(-A^3)^{-10}d(-A^3)^2(-1-A^8)(1-3A^4+4A^8-4A^{12}+4A^{16}-4A^{20}+3A^{24}-A^{28}+A^{32})\,.
\ee
which, up to $(-A^3)^w$, extra $d$ factor and substitution $A^4\to q$, is the Jones polynomial appearing from the formula in~\cite{Galakhov:2014sha} for $g=3$ and the above choice of the parameter vector.\footnote{Paper~\cite{Galakhov:2014sha} uses a slightly different convention for $q$, which is the square root of the one used here.} Let us  derive the general formula here.

Note first that using skein relation~(\ref{Udef}) it is easy to prove that
\be
\label{pretzelexemp}
R^n \ = \ A^n\, \mathbbm{1}_2 + \frac{(-A^{3})^{-n}-A^n}{d}\,\Um \ = \ A^{n}\left(\mathbbm{1}_2 + \frac{(-A^{4})^{-n}-1}{d}\,\Um\right)\,.
\ee

Let us denote $x_i=(-A^4)^{-n_i}\ = \ y_i+1$. For general genus $g$ we need to compute
\begin{multline}
\label{pretzelmatel}
\langle\Psi| A^{\sum_in_i}\bigotimes\limits_{i=0}^g \left(\mathbbm{1}_2 + \frac{(-A^{4})^{-n_i}-1}{d}\,\Um\right)|\Psi\rangle \ = \\ = \ \langle\Psi| A^{\sum_in_i}\left(\mathbbm{1}_{2^{g+1}} + \sum_i\frac{y_i}{d}\,u_{2i+1}+\sum_{i\neq j}\frac{y_iy_j}{d^2}\,u_{2i+1}u_{2j+1} + \ldots + \frac{1}{d^{g+1}}\prod_iy_iu_{2i+1}\right) |\Psi\rangle.
\end{multline}
Here only odd Temperley-Lieb generators contribute. All of them commute. 

It is easy to evaluate the matrix elements of the matrices entering the sum. 
According to the updated Kauffman's rules~(\ref{bracket2}) and~(\ref{bracket1}), they all give powers of $d$, counting the number of unlinked circles:
\be
\langle \Psi| \mathbbm{1}_{2^{g+1}}|\Psi\rangle \ = \ 
\begin{array}{c}
     \begin{tikzpicture}[thick]
         \foreach \a in {0.,0.8,...,3.2}
            \draw (0,\a) -- (1.5,\a);
         \foreach \a in {0.3,1.1,...,3.5}
            \draw (0,\a) -- (1.5,\a);
         \foreach \a in {0.3,1.1,...,2.7}
            \draw[rounded corners=3] (0,\a) -- (-0.2,\a) -- (-0.2,\a+0.5) -- (0,\a+0.5);
         \foreach \a in {0.3,1.1,...,2.7}
            \draw[rounded corners=3] (1.5,\a) -- (1.5+0.2,\a) -- (1.5+0.2,\a+0.5) -- (1.5,\a+0.5);
        \draw[rounded corners=3] (0,0) -- (-0.5,0) -- (-0.5,2.7) -- (0,2.7);
         \draw[rounded corners=3] (1.5,0) -- (2.,0) -- (2.,2.7) -- (1.5,2.7);
         \fill[white] (-0.6,0.8) rectangle (-0.4,1.1);
         \fill[white] (1.9,0.8) rectangle (2.1,1.1);
         \draw[dotted] (-0.5,1.1) -- (-0.5,0.8);
        \draw[dotted] (2,1.1) -- (2,0.8);
     \end{tikzpicture} 
\end{array} \ = \ d^{g+1}\,, \quad  
\langle \Psi| u_{2i+1}|\Psi\rangle \ = \ 
\begin{array}{c}
     \begin{tikzpicture}[thick]
         \foreach \a in {0.,0.8,...,3.2}
            \draw (0,\a) -- (1.5,\a);
         \foreach \a in {0.3,1.1,...,3.5}
            \draw (0,\a) -- (1.5,\a);
         \foreach \a in {0.3,1.1,...,2.7}
            \draw[rounded corners=3] (0,\a) -- (-0.2,\a) -- (-0.2,\a+0.5) -- (0,\a+0.5);
         \foreach \a in {0.3,1.1,...,2.7}
            \draw[rounded corners=3] (1.5,\a) -- (1.5+0.2,\a) -- (1.5+0.2,\a+0.5) -- (1.5,\a+0.5);
        \draw[rounded corners=3] (0,0) -- (-0.5,0) -- (-0.5,2.7) -- (0,2.7);
         \draw[rounded corners=3] (1.5,0) -- (2.,0) -- (2.,2.7) -- (1.5,2.7);
         \fill[white] (0.35,1.6-0.1) rectangle (1.15,1.6+0.4);
         \draw (0.35,1.6) -- (0.45,1.6) arc (-90:90:0.15) -- (0.35,1.9); 
         \draw (1.15,1.6) -- (1.05,1.6) arc (-90:-270:0.15) -- (1.15,1.9);
         \fill[white] (-0.6,0.8) rectangle (-0.4,1.1);
         \fill[white] (1.9,0.8) rectangle (2.1,1.1);
         \draw[dotted] (-0.5,1.1) -- (-0.5,0.8);
        \draw[dotted] (2,1.1) -- (2,0.8);
     \end{tikzpicture} 
\end{array} \ = \ d^{g}\,,\\ [2mm]
\langle \Psi|u_{2i_1+1}u_{2i_2+1}\cdots u_{2i_k+1}|\Psi\rangle \ = \ d^{g+1-k}\,, 
\quad k\ = \ 0,1,\ldots, g-1\,, \\ [2mm]
\langle \Psi|u_{1}u_{3}\cdots u_{2g+1}|\Psi\rangle \ = \ d^2\,. 
\ee
Hence, (\ref{pretzelmatel}) becomes
\begin{multline}
A^{\sum_in_i}\left(d^{g+1} + \sum_i y_id^{g-1} + \sum_{i\neq j}{y_iy_j}d^{g-3} + \ldots + \frac{1}{d^{g-1}}\sum_{i}\prod_{l\neq i}y_l + \frac{1}{d^{g-1}}\prod_iy_i\right) = 
\\ = \  \frac{A^{\sum_in_i}}{d^{g+1}}\left(\prod_{i=0}^{g}\left(d^2 + y_i \right)+ (d^2-1)\prod_iy_i\right)\\ = \  \frac{A^{\sum_in_i}}{d^{g+1}}\left(\prod_{i=0}^{g}\left(d^2-1 +(-A^4)^{-n_i} \right)+ (d^2-1)\prod_i((-A^4)^{-n_i}-1)\right), 
\end{multline}
which reduces to~(\ref{pretzelexemp}) for the respective choice of the parameters. Using $[3]=d^2-1$ and the conversion factors listed above we arrive at
\be
\langle\Psi| \bigotimes\limits_{i=0}^g R^{n_i}|\Psi\rangle \ = \ \frac{1}{d^{g+2}}\left(\prod\limits_{i=0}^g(1+[3](-q)^{n_i})+[3]\prod\limits_{i=0}^3(1-(-q)^{n_i})\right),
\ee
which is one of the compact general formulas for the Jones polynomials of the pretzel knots found in~\cite{Galakhov:2014sha}.

Note that the proof of the formula does not really use the specific matrix representation: the Kauffman rules are sufficient. The representation is useful for testing the general formula, as many examples, like~(\ref{pretzelexemp}) can be easily generated.

\bigskip

The method considered in these notes is a specialization of a more general tensor approach considered in~\cite{Kauffman:1987sta,Reshetikhin:1990pr,Reshetikhin:1991tc,Kauffman:2019top,Padmanabhan:2020obt}, which highlights the possibility of defining braiding operators in a certain space of states.
A similar matrix element method of computation of knot invariants exists for a unitary representation~\cite{Kaul:1991np,RamaDevi:1992np,Ramadevi:2000gq,Galakhov:2015fna,Mironov:2015aia}. In fact the unitary representation can be obtained from~(\ref{Rmatrix}) by reducing the multiplicity of its eigenvalues. However, such a reduction looses the tensor product structure of the space, where matrices act. The tensor product structure may be useful for applications. One possible class of applications is related to quantum many-body scattering problems, in which R matrix plays the role of the scattering or of the transfer matrix~\cite{Sedrakyan:2001aa,Sedrakyan:2003wh,Melnikov:2020uck}. Another class of applications is related to the topological quantum computing, in which knot diagrams correspond to quantum protocols\cite{Kitaev:1997wr,Freedman:2000rc,Nayak:2008zza,Melnikov:2017bjb}. Properties of quantum entanglement can also be addressed in the present computational framework~\cite{Kauffman:2019top,Melnikov:2018zfn,Padmanabhan:2019qed,Padmanabhan:2020obt,Melnikov:2022qyt,Melnikov:2023nzn,Melnikov:2023wwc}. Finally, since the knot invariants constructed here are also expectation values of Wilson loop operators in the $SU(2)_k$ Chern-Simons theory~\cite{Witten:1988hf},\footnote{The parameter $A$ is related to the level $k$ of the Chern-Simons theory through $A=\exp(i\pi/2(k+2))$.} it might be interesting to understand the relation of the Chern-Simons states in the present description to the conformal blocks of the associated Wess-Zumino-Witten conformal field theory~\cite{Alvarez-Gaume:1988izd}, and to the wavefunctions of the quantum Hall effect or similar topological phases of matter~\cite{Moore:1991ks}.

\bigskip

\paragraph{Acknowledgments} This work was supported in part by Simons Foundation award number 1023171-RC, grants of the Brazilian National Council for Scientific and Technological Development (CNPq) number 308580/2022-2 and 404274/2023-4 and the grant of the Serrapilheira Institute number Serra R-2012-38185.

\end{document}